\documentclass[aps,floats,epsf, twocolumn ]{revtex4}
 \usepackage{graphics}

\begin{document}

\title{Isospin of topological defects in Dirac systems}
\author{Igor F. Herbut}

\affiliation{Department of Physics, Simon Fraser University,
 Burnaby, British Columbia, Canada V5A 1S6}

\begin{abstract}
We study the Dirac quasiparticles in $d$-dimensional lattice systems of electrons
 in the presence of domain walls ($d=1$), vortices ($d=2$), or hedgehogs ($d=3$) of superconducting and/or insulating, order parameters, which appear as mass terms  in the Dirac equation. Such topological defects have been known to carry non-trivial quantum numbers such as charge and spin. Here we discuss their additional internal degree of freedom: irrespectively of the dimensionality of space and the nature of orders that support the defect, an extra
 mass-order-parameter is found to emerge in their core. Six linearly independent local orders, which close two mutually commuting three-dimensional Clifford algebras are proven to be in general possible. We show how the particle-hole symmetry restricts the defects to always carry the quantum numbers of a single effective isospin-1/2, quite independently of the values of their electric charge or true spin. Examples of this new degree of freedom in graphene and on surfaces of topological insulators are discussed.
\end{abstract}
\maketitle

\section{Introduction}

\vspace{10pt}
Massless Dirac fermions arise often in the effective theories of condensed matter. The
Fermi surface reduces to just two points in one dimension ($d$), and graphene, $\pi$-flux state, and their  generalizations
provide realizations in the dimensions two and three.\cite{hosur} p-wave and d-wave
superconductors represent yet another class of important examples.\cite{ qed3}
In the presence of some externally or dynamically generated
order parameter (OP) Dirac fermions may become gapped, with the gap
appearing as a mass-term in the Dirac Hamiltonian. Indeed, in any isotropic and time-reversal invariant Dirac system in $d=2$, such as graphene, it is known that there are precisely 36 such linearly independent masses \cite{ryu}, which stand for various components of four
insulating and four superconducting OPs \cite{dima, roy}. A pair of these masses may either commute or anticommute, with the latter ones
being particularly  interesting: they add in squares in the expression for the
total gap, and this way cooperate in the reduction of the energy of the filled Fermi sea. The examples of such pairs
 in graphene include the two phase components of any
superconducting (SC) order, or two spin components of the N\' eel OP, or of the quantum spin-Hall (QSH) topological insulator. In principle,
there exist numerous other and more exotic possibilities for such compatible OPs, in which one mass may even be insulating while the other is superconducting. \cite{ryu}

It is well-known that the domain walls and the vortex configurations in the masses of Dirac quasiparticles in $d=1$ and $d=2$ may carry electrical charge and/or spin.\cite{jackiwrebby, su, jackiw, hou}  The simplest way to understand this is on the example of insulating masses, when the electric charge represents a good quantum number: for spin-1/2 fermions there are then two zero-modes of the Dirac equation, which when fully (half) occupied yield an integrated vortex charge of one (zero) and spin zero (1/2). On the other hand, if the vortex configuration is in the SC order, for example, the particle number is not conserved, and the vortex charge is zero on  average.  Nevertheless, the existence of zero-modes of the Dirac Hamiltonian always leads to a non-trivial internal structure of the vortex, which, however, is independent of  its electrical and rotational properties. Here we discuss such an {\it additional} quantum number that generally emerges inside these topological defects, and which may be understood as their ``isospin". In particular, we investigate how the definition, value, and the physical nature of the isospin depends on the type of defect and the dimension of space in which it is embedded. Special cases of the isospin components have already been identified in graphene: a) a vortex in the easy-plane N\' eel order was claimed to contain the hard-axis N\' eel component\cite{igor4}, and b) the s-wave SC vortex was argued to develop either a density-wave (DW) \cite{ghaemi} or the QSH state \cite{igor5} in its core. We generalize here these early   examples, and in doing so, in particular, reconcile the seemingly conflicting results or refs. \cite{ghaemi} and \cite{igor5} quoted above.

 We set up and discuss the most general formulation of this problem, and find that, whereas the OPs that constitute the isospins are dependent on detail, their number and algebra is universal and {\it larger} than previously found. It is
independent of the nature of the OPs twisted into the defect, as long as their representative masses anticommute. It is also independent of the dimension of the system, if the defect is a topologically non-trivial mapping of the system's boundary onto the OP's target space. The structure we reveal is thus  identical for a domain wall in a single mass in $d=1$,\cite{jackiwrebby, su} a vortex configuration in two anticommuting masses in $d=2$,\cite{jackiw, hou, igor4, ghaemi, igor5} or a hedgehog in three mutually anticommuting masses in $d=3$.\cite{hosur,teo, herbutlu, herbut'}

  What we find is {\it six} possible linearly independent OPs in the core of the defect, which close the Clifford algebra $C(3,0)\otimes C(3,0)$,\cite{remark} but with only the averages of the linear combinations of the orders belonging to the {\it same} factor algebra as finite. This implies that the defects in question always carry a single effective isospin degree of freedom, with the quantum number of $1/2$. This is to be contrasted with the values of the real spin and the electrical charge of the defect, which both, and as we will see, even simultaneously, can average to zero. The mathematical basis of this result lies in the somewhat arcane theory of {\it real} representations of Clifford algebras.\cite{okubo} The appearance of {\it two} potentially non-zero  isospins, in particular, follows from the quternionic nature of the representations relevant to lattice fermions of spin-1/2. On the experimental side, the complementary order in the defect's core should be visible as a local mass-gap in the scanning tunneling microscopy. The existence of the isospin could lead to new ways of manipulation of the defects by the external probes that would couple to this particular degree of freedom. For example, the OP for the Kekule bond-DW order parameter couples to the pseudomagnetic field that derives from the wrinkles in the graphene sheet the same way the usual electric charge density couples to the real magnetic field. The possibility that defects in one OP carry with them quantum numbers of another is also suggestive of an exotic quantum phase transition from their proliferation. \cite{senthil}

   A reconsideration  of the defects in the insulating OPs becomes particularly interesting in light of our results, which imply that one of the potential isospins consists of a single insulating and two superconducting components. The average electric charge of such a defect is a continuous variable between zero and one. The chemical potential in this case polarizes the isospin in the direction of the insulating component. It therefore causes the isospin to {\it precess}, which translates into a periodic time-dependence of the local {\it superconducting} phase.

  This paper is organized as follows. In the next section we define the Dirac-Nambu Hamiltonian in two dimensions, and discuss the consequences of the particle-hole symmetry, particularly of its antilinearity. The vortex Hamiltonian is then studied in Sec. III. In particular, in this section we show that the maximal Clifford algebra that contains the two gamma- and two mass-matrices appearing in the vortex Hamiltonian is closely related to $C(2,5)$.
  A generalization to other dimensions and the connection to Wess-Zumino-Witten term is briefly described in Sec. IV. The way the isospin develops a non-zero expectation value from the zero-modes, and its further unitary evolution in time are considered in Sec. V. Some examples are provided in Sec. VI. The connection between the isospin on one side, and the usual spin and the electric charge on the other is discussed in the closing section, where we also summarize the main results. In the Appendices we provide the derivation of the Majorana representation used throughout the paper
  and give an elementary discussion of the crucial mathematical fact about the Clifford algebra  $C(2,5)$, that is its quaternionic nature.

  \section{Dirac-Nambu Hamiltonian}

  To derive the isospin we take $d=2$ and generalize to other spatial dimensions afterwards. Assume a space- and time-inversion invariant system of spin-1/2 fermions on a lattice, with the  Dirac point/points in its band structure, such as graphene. The Dirac Hamiltonian for all low-energy excitations near the Dirac points will then have the standard form
\begin{equation}
H_0 = \alpha_1 k_1 + \alpha_2 k_2 +O(k^2),
\end{equation}
where $\alpha_i$, $i=1,2$ are {\it eight-dimensional} Hermitian matrices satisfying the anticommutation relation $\{ \alpha_i, \alpha_j \} = 2 \delta_{ij}$, and $\vec{k}=-i\nabla $ is the usual momentum operator. The dimension eight may be understood as arising, in graphene, for example, from two sublattices, two valleys, and two spin projections, but it is in fact required more generally  by the time reversal invariance, as discussed in detail in ref. \cite{herbut'}. To give a unified description of both the insulating and the superconducting mass-gaps one must double the number of fermionic  components, include both particles  and holes into the Dirac spinor, and consider the {\it sixteen-dimensional} massive ``Dirac-Nambu" Hamiltonian
\begin{equation}
H= H_k + m M
\end{equation}
where, as usual, the kinetic energy part is
\begin{equation}
H_k = H_0\oplus (-H_0 ^*)\equiv \Gamma_i k_i,
\end{equation}
and the matrices $M$ and $\Gamma_i$, $i=1,2$ are Hermitian. The symbol ``$\oplus$" stands for an orthogonal sum. To open a mass-gap we require that $\{ M, \Gamma_i\} =0$ and $M^2=1$, so that the spectrum of $H$ is the usual relativistic $ \pm \sqrt{k^2 + m^2 }$ for $m$ uniform.

The Dirac-Nambu Hamiltonian $H$ is by construction particle-hole symmetric: it anticommutes with the antilinear operator $I_{ph}=(\sigma_1 \otimes I)K$, where $I$ is the eight-dimensional unit matrix, $\vec{\sigma}$ are the Pauli matrices, and $K$ is the complex conjugation.  Since $I_{ph} ^2 = 1$, a representation exists in which $I_{ph} =K$ \cite{messiah, altland}, and $H$ is imaginary. Explicit derivation is provided in  Appendix 1. Such a representation could be called ``Majorana", and we will assume it hereafter. In the Majorana representation any mass-matrix $M$ is also imaginary. Since, on the other hand, in the coordinate representation the momentum operator $\vec{k}$ is already imaginary, the two sixteen-dimensional matrices  $\Gamma_i$, $i=1,2$ in the kinetic energy are {\it real}. The separation of the matrices in the kinetic and mass terms into real and imaginary essentially determines the internal structure of the vortex, as will be discussed shortly.

\begin{table}[t]

\begin{tabular}{c || c c c c c c c c c r }
p,q & 0& 1 & 2 &3&4&5&6&7&8 \\
 \hline
 \hline
0&  &\em{2}&\bf{4}&\bf{4}&\bf{8}&\em{8}&8&8&16 \\
1& 1&2&\em{4}&\bf{8}&\bf{8}&\bf{16}&\em{16}&16&16 \\
2& 2&2&4&\em{8}&\bf{16}&\bf{16}&\bf{32}&\em{32}&32 \\
3& \em{4}&4&4&8&\em{16}&\bf{32}&\bf{32}&\bf{64}&\em{64} \\
4& \bf{8}&\em{8}&8&8&16&\em{32}&\bf{64}&\bf{64}&\bf{128}\\
5& \bf{8}&\bf{16}&\em{16}&16&16&32&\em{64}&\bf{128}&\bf{128}& \\
6& \bf{16}&\bf{16}&\bf{32}&\em{32}&32&32&64&\em{128}&\bf{256} \\
7& \em{16}&\bf{32}&\bf{32}&\bf{64}&\em{64}&64&64&128&\em{256} \\
8& 16&\em{32}&\bf{64}&\bf{64}&\bf{128}&\em{128}&128&128&256 \\

\end{tabular}

\caption[]{The dimensions of irreducible real representations of Clifford algebras $C(p,q)$, with $p$ labeling the rows, and $q$ labeling the columns. The normal font denotes the ``normal" representations (with one Casimir operator), the emphasized font denotes the ``almost complex" representations (with two Casimirs), and the bold font denotes the ``quaternionic" representations (with four).}

\end{table}

\section{Vortex and its internal structure}

Let us then assume a real-space vortex configuration in the mass term,
  \begin{equation}
M (\vec{x}) = M_V = M_1 \cos\theta  + M_2 \sin\theta ,
  \end{equation}
with $(r,\theta)$ as the polar coordinates, and $m=m(r)$ in Eq. 2 and arbitrary.  The only requirements are that the two Hermitian matrices $M_1$ and $M_2$ anticommute, $\{M_i, M_j \}= 2 \delta_{ij}$, again be imaginary, and anticommuting with the kinetic energy, i.e. $\{M_i, \Gamma_j\}= 0$.

The spectrum of the Hamiltonian $H_k + m M_V$  now contains {\it four} zero-energy states. This follows from noticing that the matrices $\Gamma_i$ and $M_i$, $i=1,2$ which appear in the Hamiltonian form a Clifford algebra $C(4,0)$, which, as is well known, has a unique four-dimensional irreducible representation.\cite{schweber} The Dirac-Nambu Hamiltonian can therefore be transformed into an orthogonal sum of four identical copies of the
 Jackiw-Rossi Hamiltonian, which contains one zero-mode in its spectrum.\cite{jackiw} Any operator that anticommutes with the Hamiltonian may, depending on its  occupation, pick up an expectation value precisely from those zero-energy states \cite{igor4, ghaemi, igor5}. To understand the internal structure of the vortex we therefore need to find the imaginary Hermitian matrices $X$ that satisfy
\begin{equation}
\{X,\Gamma_i\} = \{X, M_i \} =0,
\end{equation}
$i=1,2$, and $X^2 =1$.

In the Majorana representation the problem is solved by invoking some properties of the real representations of the Clifford algebras.  To see how this algebraic structure emerges we first ask how many such {\it mutually anticommuting} mass-matrices $X$ exist. The answer is contained in Table I, where we extracted and summarized the required information from ref. \cite{okubo}, both on the dimensionality and the type of the irreducible {\it real} representations of the Clifford algebras $C(p,q)$. Since $\Gamma_i$, $i M_i$, and $iX$ are all real, and $\Gamma_i ^2 =+1$ and $(i M_i)^2 = (i X )^2 =-1$, we need to determine the maximal value of $q$ so that for $ p\geq 2$ the dimensionality of the real representation of $C(p,q)$ is sixteen. An examination of the table immediately implies that the maximal $q=5$, as observed already in the ref. \cite{ryu}. There are thus {\it three} additional mutually anticommuting masses: $X=M_3, M_4, M_5$, so that the set $( \Gamma_i, i M_j )$, $i=1,2$, $j=1,2,3,4,5$ provides a real irreducible sixteen-dimensional representation of $C(2,5)$.

This is not all, however. The essential feature of the real representation of $C(2,5)$ is that it is {\it quaternionic}: besides the usual unit matrix, there exist three non-trivial linearly independent sixteen dimensional matrices which commute with the whole representation.
 This facilitates the construction of three {\it additional} solutions: $X=M' _3, M' _4, M' _5$. While mutually anticommuting ($\{ M' _i, M' _j \}= 2\delta_{ij}$), these solutions {\it commmute} with the first three: $[M_i, M' _j ] = 0 $, for $i,j=3,4,5$. The mathematical details of this construction are provided in Appendix 2.

   All possible ``five-tuplets" of anticommuting masses which respect the particle-hole symmetry have been listed with the help of a computer in ref. \cite{ryu}.
From this list it is evident, under close inspection, that each pair of symbols, irrespective of what physical OPs they may represent, can be found in exactly {\it two} different five-tuplets. Any set of three, or four, symbols, on the other hand, belongs to a unique five-tuplet. All of these facts follow from the representation theory as explained above.

\begin{figure}[t]
{\centering\resizebox*{80mm}{!}{\includegraphics{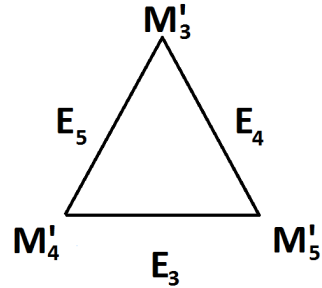}}
\par} \caption[] {The three mutually anticommuting OPs (the first isospin's components) in the defect's core.  The OP at each corner commutes with the symmetry operator at the opposite side of the triangle, which rotates the remaining two OPs into each other. There also exists the second triangle of OPs, with $M_i ' \rightarrow M_i$ (the second isospin's components) and $E_i \rightarrow E_i '$, which is not shown. All the matrices are imaginary, and all the matrices of the first triangle commute with those of the second.}
\end{figure}

\section{Other dimensions and Wess-Zumino-Witten term}

The {\it six} masses that we found to anticommute with the Dirac-Nambu Hamiltonian in $d=2$ with a mass-vortex close the algebra $C(3,0) \otimes C(3,0)$, with the four-dimensional Hilbert space of zero-energy states ${\cal H}_0$ as its invariant subspace. Table I implies, however, that this result is independent
 of the spatial dimensions $d$: since the dimensions of the matrices appearing in the Dirac-Nambu Hamiltonian \cite{herbutlu} are $2^{d+2}$, $d$ real $\Gamma$-matrices and  $d+3$ imaginary masses, upon multiplication of the latter by $i$, form a real representation of $C(d, d+3)$, which is always quaternionic. Assuming the topological defect in $d$ of the anticommuting masses again leaves the algebra  of potential core order parameters $C(3,0)\otimes C(3,0)$.  The general result can then be summarized as in Fig. 1.

In any dimension $d$, the number of mutually anticommuting masses, $d+3$, is precisely right for the emergence of the Wess-Zumino-Witten term upon the integration over fermions.\cite{ghaemi, jaroszewicz, pasha, tanaka} The Berry phase of the three OPs that can form in the core of the defect of the remaining $d$ masses will thus match the one of a spin-1/2-like object, as discussed in \cite{tanaka}, and in accord with our assignment of the isospin's quantum number.

\section{Two isospins and zero-mode subspace}

\subsection{State of isospin}

Since the zero-mode subspace ${ \cal H}_0$ is four-dimensional, we may choose a basis in which the first three core OPs  ($M_i$) are represented by $S_{1i}=\sigma_i \otimes \sigma_0$, and the second three ($M'_i$)  by $S_{2i}=\sigma_0\otimes \sigma_i$ ($i=3,4,5$). It would then seem that the averages of both, or possibly none, of the two isospins could be finite in the ground state. That this however, is impossible, is guaranteed by the antilinearity of the particle-hole symmetry in a simple but rather interesting way. Consider a perturbed Dirac-Nambu Hamiltonian
\begin{equation}
H= H_k + m M_V + \epsilon O,
\end{equation}
where $O$ is an operator so far neglected, which derives, for example, from the underlying lattice, magnetic field, finite chemical potential, or a similar perturbation. For small enough parameter $\epsilon$  we may neglect the mixing of the zero-energy and finite-energy modes, and to the leading order consider only the resulting split of the zero-modes. We now show that the result of {\it any} such splitting can only be that either  $\langle \vec{S}_1 \rangle\neq 0$ and $\langle \vec{S}_2 \rangle= 0$ or vice versa.

Since in the Majorana representation the matrix  $O$ for {\it any} perturbation is also imaginary, $O$ anticommutes with some antilinear operator $A$ that represents the particle-hole symmetry within the zero-mode subspace ${ \cal H}_0$. Introduce the basis in ${ \cal H}_0$:
$\{ |\pm,\pm \rangle \}$, with  $S_{\alpha 3} |s_1, s_2 \rangle  = s_\alpha |s_1,s_2 \rangle$,
where $s_\alpha = \pm$, and $\alpha =1,2$ labels the isospin. If $|\psi\rangle$ is an eigenvector of the matrix representing $O$ in ${\cal H}_0$, its eigenvalue has the opposite sign of the eigenvalue of the eigenstate $A|\psi\rangle$. The two states are thus orthogonal.  Since $\vec{S}_\alpha$, $\alpha =1,2$ are also odd under particle-hole symmetry, it follows that within ${ \cal H}_0$ $A$ is formally equivalent to the time reversal of spin. In general then, if
\begin{equation}
|\psi \rangle = a |+,+\rangle + b |+,-\rangle + c |-,+\rangle +d  |-,-\rangle,
\end{equation}
its particle-hole partner is
\begin{equation}
A |\psi \rangle = -a^* |-,- \rangle + b^* |-,+\rangle + c^* |+,-\rangle - d^*  |+,+\rangle,
\end{equation}
where $A|\pm \rangle = \pm i |\mp\rangle$. The orthogonality of the two states implies that
 \begin{equation}
 \frac{c}{a}=\frac{d}{b},
 \end{equation}
 which is precisely what  disentangles $| \psi\rangle$ into a {\it product state}:
\begin{equation}
|\psi \rangle = (|+\rangle_1 + \frac{c}{a} |-\rangle_1 ) ( a|+\rangle_2 + b|-\rangle_2)= |\vec{n}\rangle_1 |\vec{m}\rangle_2,
\end{equation}
where, in the usual notation,
\begin{equation}
\vec{t}\cdot\vec{S}_\alpha  |\vec{t}\rangle_\alpha = |\vec{t}\rangle_\alpha
\end{equation}
for $\alpha=1,2$, and $\vec{t}^2 =1$.

If the state $|\psi \rangle= |\vec{n}\rangle_1 |\vec{m}\rangle_2$ corresponds to a negative eigenvalue of $O$, the state  $ A |\psi \rangle=|-\vec{n}\rangle_1 |-\vec{m}\rangle_2$ corresponds to a positive eigenvalue. Since the other two eigenstates are also the product states, and need to be orthogonal to the first two, there are just two options: it is either that the state $ |-\vec{n}\rangle_1 |\vec{m}\rangle_2$ has then the negative energy, and $ |\vec{n}\rangle_1 |-\vec{m}\rangle_2$ positive, or vice versa. In the former case, $\langle \vec{S}_2 \rangle =\vec{m} $,  and $\langle \vec{S}_1 \rangle = 0$. The state of the isospins may thus be understood as a {\it mixture}, given by the statistical operator (i.e. the density matrix)
\begin{equation}
\rho = \frac{1}{2} \sigma_0 \otimes |\vec{m}\rangle \langle \vec{m} |.
\end{equation}
In the latter case, of course, $\langle \vec{S}_1 \rangle =\vec{n}$, and $\langle \vec{S}_2 \rangle = 0$. The state of the two isospins is a product of the maximal mixture ($1/2 \sigma_0$) for one, and the pure state ($|\vec{m}\rangle \langle \vec{m} |$) for the other isospin.

\subsection{Time evolution}

 The unitary evolution cannot exchange the two isospins: if $\langle \vec{S}_1 \rangle = 0$ $t=0$, for example, it remains zero at $t>0$. To see this, assume that the isospin's state initially is $\rho (0) = (1/2) \sigma_0 \otimes \rho_2 (0)$ with a general statistical operator $\rho_2 (0)$. It evolves according to the Schr\"
{o}dinger equation
  \begin{equation}
  i\dot{\rho}(t) = [O, \rho(t)].
  \end{equation}
 The particle-hole symmetry implies that in general the perturbation, which may be time-dependent, within the zero-mode subspace has the form
  \begin{equation}
  O= f(t) P_1 (t) \otimes \sigma_0 + h(t) \sigma_0 \otimes P_2 (t),
  \end{equation}
 where $P_i (t)= | \vec{r}_i (t) \rangle \langle \vec{r}_i (t) | $ are time-dependent projectors on some one-dimensional subspaces of the first and the second isospin, and $f(t)$ and $h(t)$ are arbitrary functions of time. The time-dependent state is therefore $\rho (t) = (1/2) \sigma_0 \otimes \rho_2 (t)$, with
 \begin{equation}
  i\dot{\rho}_2(t) = h(t) [P_2 (t), \rho_2(t)],
  \end{equation}
and, as evident,  $\langle \vec{S}_1 \rangle = 0$ at all times.

The topological defects thus ultimately carry with them {\it a single} non-trivial isospin-1/2, of the physical size of the spatial extent of the zero-modes, which is $\sim 1/m(\infty)$. The OPs forming the components of the isospin depend on the system. Let us consider some examples next, starting from the most familiar.

\section{Examples}

\subsection{Superconducting vortices}

 Let us consider graphene ($d=2$), as an example. If the vortex configuration is in the two components of the s-wave SC, core's isospin components are the charge-DW and the two singlet bond-DWs,\cite{ghaemi} or the three components of the QSH state \cite{igor5}. The real spin in the former case is zero. If the superconductor is a spin singlet, as in this example, one of the isospins must be a vector under rotations, since all three spin-rotation generators commute with the Hamiltonian, and will appear as the Casimir operators $E_3, E_4, E_5$ in Fig. 1, for one of the isospins. Obviously, in this case it is the QSH state that carries the (true) spin 1/2, which is why it is favored by the Zeeman term in presence of the magnetic field.\cite{igor5} In general, if the matrices $M_1$ and $M_2$ in Eq. (4) represent two phase components of some superconducting order,\cite{roy} all six core's OPs are insulating. The electric charge is not a good quantum number, but on average it always vanishes, independently of the state of isospin. The lattice itself, for example, favors the bond-DW in the core: this is thus an example of a vortex with vanishing both (average) charge and true spin, and with only the {\it isospin} left as finite.

The situation changes if one twists two spin components of a triplet SC; $x$ and $y$ components of the f-wave SC in graphene, \cite{roy, honerkamp} for example. Each potential  isospin then has a single insulating and two superconducting components: 1) CDW, s-wave SC, $z$ component of the f-wave SC, and 2) $z$ components of N\' eel, and two Kekule triplet superconducting states. In either case the vortex core may stay superconducting by simply changing the type of the superconducting state, i. e. by forming a {\it texture}. Neither electric charge nor spin are good quantum numbers in this case, but the isospin is well-defined.

\subsection{Insulating vortex}

 Next, consider a vortex configuration in two insulating OPs, when the particle number, and hence the electric charge, is a good quantum number. For instance, the spin-singlet (Kekule) bond-DW \cite{hou} is favored by the coupling to in-plane optical phonon at the Dirac point in graphene, and has been proposed as a candidate for the OP in the magnetic field.\cite{nomura, mudry, herbutfield}  The isospin components in the vortex configuration in this case are, 1) the three components of the N\' eel OP, or 2) the charge-DW and the two phase components of the s-wave SC.  The vortex is (spin) rotational symmetric, and the spin is 1/2 and zero, respectively. The electric charge for the first isospin is zero.

 Before discussing the electric charge of the second isospin, let us consider the vortex formed out of the $x$ and $y$ components of the N\' eel OP, favored by the Hubbard on-site repulsion.\cite{herbutU} One set of isospin components contains the $z$-component of the same N\' eel order \cite{igor4} and two spin-singlets bond-DWs, and the other consists of the $z$-component of the QSH state and the two phase components of the $z$-component of the f-wave triplet SC \cite{roy}. Spin is not a good quantum number.

 In both examples introduced above, one isospin (call it $\vec{S}_1$) is purely insulating, whereas the second one ($\vec{S}_2$) has one insulating and two superconducting components. It is easy to see that this is in general true; the particle number is conserved, and thus the number operator must be one of the Casimir operators, let's say $E_3$ in Fig. 1. The masses $M_4'$ and $M_5'$ must then stand for two phase components of some superconducting order, rotated into each other by a transformation generated by the number operator.

 Let us further assume that in the presence of some perturbation we found $\langle \vec{S}_2 \rangle = \vec{n}$, where $\vec{n}=(\sin\theta \cos\phi, \sin\theta\sin\phi,\cos\theta)$,  with $S_{23} = \sigma_0\otimes \sigma_3$ as the insulating component of the second isospin. The particle number operator within ${\cal H}_0$ is then also $N=\sigma_0\otimes \sigma_3$, and the average of the vortex electric charge is $\langle N \rangle = \langle \vec{n}|\sigma_3|\vec{n}\rangle = \cos\theta$. Of course, in this example the charge will be sharp only if $\theta=0$, which will be the case if the perturbation in question is the chemical potential $\mu N$, for instance. In the ground state then $\theta=0$, and the charge is back to unity. One can imagine, however, that a proximity to another superconductor could ``polarize" the core so that $\theta\neq 0$. At a finite chemical potential the  superconducting phase will then oscillate as $\phi(t) = \phi (0) + \mu t$, whereas $\theta$ is constant, just as in the precession of an ordinary spin.

 \subsection{Topological insulator}

Our results do not, however, readily apply to a superconducting vortex on an isolated surface of a topological insulator, if the original fermion has only two components. After the particle-hole doubling the dimension of the matrices involved becomes four, and Table I implies that for $p=2$ the maximal value of $q$ is only two. The set of the core's order parameters is simply empty, in accord with the single zero-energy Majorana state being pinned to the chemical potential.\cite{fu} If we consider a thin film geometry with two possibly coupled surfaces, however, the dimension of the Dirac-Nambu Hamiltonian doubles to eight, and the maximal value of $q$ increases to $q=3$. The situation still differs qualitatively from the one we studied hitherto in that there is now a single isospin in the core, and with  the discrete Ising, and not Heisenberg symmetry. If the vortex is in the s-wave SC, for example, one finds that the core's Ising isospin is the time-reversal-invariant exciton of ref. \cite{babak}.

\section{Summary and discussion}

  To conclude, we showed that some typical topological defects in the presence of Dirac fermions, besides their usual electric charge and true spin, which, depending on the defect, may or may not be finite, always possess a non-zero isospin. The components of the isospin are the local masses, i. e. order parameters, that develop a finite expectation value in the center of the defects. While in principle two different such isospins are always possible, only one can have a non-trivial quantum number in a given physical situation. This new degree of freedom couples to different external probes, and in principle could be manipulated similarly to the electron spin.

  The connection between the isospin, discussed here, and the electrical charge and the true spin of the vortex now becomes transparent. A particular split of the zero-modes favored by the perturbations to the Dirac Hamiltonian will result in a component of one of the two possible isospins developing a non-zero average value; say, for instance, $\langle \Psi^\dagger M_3 ' \Psi \rangle\neq 0$. The vortex is thus by this mechanism {\it spontaneously} turned into a meron. There is a unique non-trivial symmetry generator that commutes both with the vortex Hamiltonian, and with the matrix $M_3 '$: it is $E_3$ in Fig. 1. The commutation relations between the six masses $M_i$ and $M_i '$ and the six Casimir operators $E_i$ and $E_i '$ imply, however, that {\it within the zero-mode subspace}, $E_3$ and $M_3 '$ are represented by the same matrix, and therefore it is also $\langle \Psi^\dagger E_3  \Psi \rangle\neq 0$. The matrix $E_3$ can be the number operator, in which case the latter average would be proportional to the density of the electric charge. It can also be the generator of rotations of the electron spin, in which case the  vortex will have the spin 1/2. The examples of both were discussed in the previous section. But $E_3$ can also be {\it neither}. For a singlet-SC vortex on graphene's honeycomb lattice, for example, if the non-zero isospin is Kekule bond-DW, $E_3$ will be one of the generators of valley rotations.

\section{Acknowledgement}

This work was supported by the NSERC of Canada. I am grateful to D.-H. Lee, C.-K. Lu and B. Roy for useful discussions, and The Princeton Center for Theoretical Science for hospitality.

\section{Appendix 1: Majorana representation}

Particle-hole symmetry dictates that the mass matrix $M$ satisfies the constraint:
\begin{equation}
M= - (\sigma_1 \otimes I) M^\top (\sigma_1 \otimes I)
\end{equation}
where $I$  is the eight-dimensional unit matrix, and $\vec{\sigma}$ are the Pauli matrices. The constraint is the expression of the Fermi statistics, and it implies that the bilinear $\Psi^\dagger M \Psi  $ is finite.

 The determination of all possible masses is therefore made non-trivial by the constraint. Luckily, it can be simplified by a unitary transformation $\tilde{M} = U M U^\dagger $, so that $\tilde{M}$ would satisfy $\tilde{M}= - \tilde{M} ^\top $. Here $U=U_2 \otimes I$, with the two-dimensional unitary matrix $U_2$ satisfying
\begin{equation}
\sigma_0 = U_2 \sigma_1 U_2 ^\top.
\end{equation}
The general solution is then
\begin{equation}
U_2 = (\pm i)^{1/2} e^{ i \frac{\pi}{4} \sigma_3}  e^{ i \frac{\pi}{4} \sigma_2} e^{ i (\frac{\pi}{4} -\phi) \sigma_3}.
\end{equation}
$\phi=\pi/4$, for example, yields the matrix of ref. \cite{altland}. Since the matrix $\tilde{M}$ is also Hermitian, it follows that $\tilde{M} = - \tilde{M}^*$,
i. e. that it is purely {\it imaginary}. Also, $\{ \tilde{M}, \tilde{H}_k \}  = 0$,  where $\tilde{H}_k = U H_k U^\dagger$.

Writing $\alpha_i = Re(\alpha)_i + i Im(\alpha)_i$, the kinetic energy of the Dirac-Nambu Hamiltonian takes the form
\begin{equation}
H_k = ( \sigma_0 \otimes Re(\alpha)_i + i \sigma_3 \otimes Im(\alpha)_i) k_i\equiv \Gamma_i k_i.
\end{equation}
Since, on the other hand, $U_2 \sigma_3 U_2^\dagger = \sigma_2$, after the unitary transformation the same kinetic energy becomes
\begin{equation}
\tilde{H}_k =  ( \sigma_0 \otimes Re(\alpha)_i + i \sigma_2 \otimes Im(\alpha)_i) k_i \equiv \tilde{\Gamma}_i k_i,
\end{equation}
with the matrices $\tilde{\Gamma}_i$ being manifestly {\it real}.

\section{Appendix 2: quaternionic representation of C(2,5)}

The crucial feature of the real sixteen-dimensional representation of the Clifford algebra $C(2,5)$ is that it is {\it quaternionic}:
besides the unit matrix, there exist {\it three} additional real Casimir matrices, $K_i$, $i=3,4,5$, which obey the quaternion algebra,
\begin{equation}
K_i K_j = -\delta_{ij} + \epsilon_{ijk} K_k,
\end{equation}
and which commute with all seven matrices $\{ \Gamma_i, iM_j \}$, $i=1,2$, $j=1,2,3,4,5$.
Defining then the imaginary matrices as $E_i = i K_i$, $i=3,4,5$, we see that they mutually anticommute, $\{E _i, E_j \} = 2\delta_{ij}$, and close the $SU(2)$ Lie algebra. Their existence enables one to construct additional solutions for the matrices $X$ as follows. Consider,
\begin{equation}
M' _i = E_i \Gamma_1 \Gamma_2 M_1 M_2
\end{equation}
and $i=3,4,5$. Obviously, $\{ M' _i, \Gamma_j \} =   \{ M' _i, M _j \}  =0$
for $j=1,2$. $M'_i$, $i=3,4,5$ are both imaginary and Hermitian, and $\{ M' _i, M ' _j\} = 2 \delta_{ij}$. The set $\{ M_i, M' _j \}$, $i=1,2$, $j=3,4,5$ is then a five-tuplet of mutually anticommuting masses, {\it different} from $\{ M_i, M_j \}$. We also see that $[M_i, M' _j ] = 0 $ for $i,j=3,4,5$, and, by Eq. 1,
\begin{equation}
i E_3  M' _4 = M' _5,
\end{equation}
 together with cyclic permutations of the indices. $M' _i$, $i=3,4,5$ therefore form a vector under transformations generated by $E_i$. Of course, the real  representation of $C(2,5)$ provided by $\{ \Gamma_i, i M_j , i M' _k\}$, $i,j=1,2$, $k=3,4,5$, is also quaternionic, with its own set of Casimir operators $E' _i$, $i=3,4,5$, so that it is also true that
\begin{equation}
M _i = E' _i \Gamma_1 \Gamma_2 M_1 M_2.
\end{equation}

The origin of the non-trivial Casimir operators is easy to understand. The maximal number of complex sixteen-dimensional mutually anticommuting matrices is {\it nine}. If they are all to square to $+1$, five of them, call them $R_i$, $i=1,2,3,4,5$ can be real,  and the other four, $I_i$, $i=1,2,3,4$,  imaginary. This is a direct generalization from the case of two-dimensional Pauli matrices, and is in accord with the normal nature of the real representations of $C(q+1,q)$ in the Table I in the text. We may take that, for example,
\begin{equation}
R_5 =  \prod_{i=1} ^4 R_i I_i.
\end{equation}

 Form then the subset of six matrices $\{ R_i, I_k\}$, $i=1,2$, and $k=1,2,3,4$. Obviously, we can construct an unique additional imaginary matrix out of the three omitted real matrices as
  \begin{equation}
  I_5 = i R_3 R_4 R_5,
  \end{equation}
which then anticommutes with all six matrices of the subset. The three imaginary Casimir operators are then simply
\begin{equation}
E_i = \frac{i}{2} e_{ijk} R_j R_k,
\end{equation}
where $i,j,k= 3,4,5$ in the last equation. Generalization to other dimensions is straightforward. One can similarly understand the ``almost complex" real representations of $C(p, p+1)$, with two Casimir operators, which is relevant for the vortices on two surfaces of topological insulator thin films discussed in the text, or for spinless fermions on honeycomb lattice.


\begin{thebibliography}{99}
\bibitem{remark} $C(p,q)$ is defined as a set of $p+q$ mutually anticommuting
 generators, $p$ ($q$) of which square to $+1$ ($-1$).
\end{thebibliography}
\end{document}